\documentclass[12pt]{iopart}
\usepackage{iopams}  
\usepackage[breaklinks=true,colorlinks=true,linkcolor=blue,urlcolor=blue,citecolor=blue]{hyperref}
\usepackage{rotating}
\usepackage{epsfig}
\usepackage{graphicx}
\usepackage{epstopdf}

\begin{document}

\pagenumbering{arabic}

\title{Electronic structure of the Mo$_{1-x}$Re$_x$ alloys studied through resonant photoemission spectroscopy} 

\author{Shyam Sundar$^{1,3}$, Soma Banik$^{2}$, L S Sharath Chandra$^{3}$, M K Chattopadhyay$^{1,3}$, Tapas Ganguli$^{1,2}$, G S Lodha$^{1}$, Sudhir K Pandey$^{4}$, D M Phase$^{5}$ and S B Roy$^{1,3}$}
\address{$^1$Homi Bhabha National Institute at RRCAT, Indore, Madhya Pradesh 452013, India}
\address{$^2$Materials Research Lab, Indus Synchrotrons Utilization Division, Raja Ramanna Centre for Advanced Technology, Indore, Madhya Pradesh 452013, India. }
\address{$^3$Magnetic and Superconducting Materials Section, Raja Ramanna Centre for Advanced Technology, Indore, Madhya Pradesh 452013, India.}
\address{$^4$School of Engineering, Indian Institute of Technology, Mandi, Kamand, Himachal Pradesh-175005, India.}
\address{$^5$UGC-DAE Consortium for Scientific Research, Khandwa Road, Indore, Madhya Pradesh-452001, India.}

\begin{abstract} 

We have studied the electronic structure of Mo rich Mo$_{1-x}$Re$_x$ alloys (0$\leq$ x $\leq$0.4) using valence band photoemission spectroscopy in the photon energy range 23-70~eV and density of states calculations. Comparison of the photoemission spectra with the density of states calculations suggests that with respect to the Fermi level $E_F$, the $d$ states lie mostly in the range 0 to -6~eV binding energy whereas $s$ states lie in the range -4 to -10 eV binding energy. We have observed two resonances in the photoemission spectra of each sample, one at about 35~eV photon energy and other at about 45~eV photon energy. Our analysis suggest that the resonance at 35~eV photon energy is related to the Mo $4p$-$5s$ transition and the resonance at 45~eV photon energy is related to the contribution from both the Mo $4p$-$4d$ transition (threshold: 42~eV) and Re $5p$-$5d$ transition (threshold: 46~eV). In the CIS plot, the resonance at 35~eV incident photon energy for binding energy features in the range of $E_F$ (B.E. = 0) to -5~eV becomes progressively less prominent with the increasing Re concentration $x$ and vanishes for $x>$ 0.2. The difference plots obtained by subtracting the valence band photoemission spectrum of Mo from that of Mo$_{1-x}$Re$_x$ alloys, measured at 47 eV photon energy, reveal that the Re $d$ like states appear near $E_F$ when Re is alloyed with Mo. These results indicate that interband $s$-$d$ interaction, which is weak in Mo, increases with the increasing $x$ and influences the nature of superconductivity in the alloys with higher $x$.
\end{abstract}

%\submitto{\JPCM}

%\keywords{photoemission spectroscopy, Electronic structure, Transition metal binary alloy superconductors}
\vspace{2pc}

\maketitle

\section{Introduction}

Discovered in 1960s, the binary Mo-Re alloys are interesting superconductors having transition temperature ($T_c$) about an order of magnitude higher than their constituent elements \cite{hul61, mor63, von82}. Experimental studies done in mid 1980s and early 1990s suggested the possibility of Re influenced softening of the phonon spectrum \cite{shu86} and a change in the electron density of states (DOS) \cite{tul93} in the Mo$_{1-x}$Re$_x$ alloys. But these phenomena could not completely explain the large non-monotonic  enhancement of $T_C$ in the Mo$_{1-x}$Re$_x$ alloys. The $T_c$ in the Mo$_{1-x}$Re$_x$ alloys increases slowly from 0.90~K for Mo to about 3~K for $x$ = 0.10 and then rises sharply to about 7~K for $x$ = 0.15 \cite{ign80, shy15a}. With further increase in $x$, the $T_c$ increases linearly to about 12.6~K for $x$ = 0.40 \cite{ign80, shy15a}. The composition range in which the $T_c$ increases sharply in the Mo$_{1-x}$Re$_x$ alloys, corresponds to the same composition range where the existence of two electronic topological transitions (ETT) have been reported for the critical concentrations $x_{c1}$ = 0.05 and $x_{c2}$ = 0.11 \cite{vel86, gor91, sko94, ign02, ign07, sko98, oka13}. The ETT is associated with the appearance or the disappearance of pockets of Fermi surface when an external parameter such as composition, pressure, and/or magnetic field is varied \cite{bru94}. The coefficient of the thermoelectric power $\alpha/T$ in the zero temperature limit shows a giant enhancement around $x_{C2}$ = 0.11 \cite{vel86, ign02, ign07}. The $T_c$ of the Mo$_{1-x}$Re$_x$ alloys with $x > x_{c2}$ is reported to oscillate with pressure \cite{ign80, ign07}. The oscillations in the temperature derivative of $\alpha/T$ and resistivity were also observed to be maximum at $x_{c1}$ = 0.05 and $x_{c2}$ = 0.11 \cite{ign07}. These oscillations were predicted to be due to the localization of electrons at the newly appeared Fermi pockets \cite{ign07}. Our recent band structure calculation shows that there is substantial changes in the DOS at the Fermi level $E_F$ for $x > x_{c2}$ \cite{shy15a}. The X –ray photoelectron spectroscopy revealed that the rigid band model is not applicable in the case of the Mo$_{1-x}$Re$_x$ alloys, and the changes in the spectra as a function of $x$ was assigned to the ETT \cite{yar86}. The direct evidence of this ETT has been provided recently by the Okada et. al., with the help of angle resolved photoemission spectroscopy along the H-N direction of the Brillouin zone \cite{oka13}. However, their studies could not establish any relation between the ETT and the superconductivity in Mo$_{1-x}$Re$_x$ alloys.  

In our recent study, we have found the evidence of multiband effects in the temperature dependence of heat capacity and the superfluid density (estimated from the temperature dependence of lower critical field H$_{c1}$) in the Mo$_{1-x}$Re$_x$ alloys with $x$ = 0.25 and 0.40 \cite{shy15b}. Our detailed study also reveals that the multiband superconductivity appears above $x>$ 0.11 which is the critical composition corresponding to the ETT \cite{shy15c}. From Helium channeling experiments \cite{kum79} on Mo$_{1-x}$Re$_x$ alloys, Dikiy et. al., concluded that the new electron sheet that is formed above $x > x_{c2}$ differ from the main sheets in terms of lower velocity and higher effective mass \cite{dik06}. Resonant photoemission spectroscopy (RPES) has been shown to be a powerful tool for determination of partial density of states in various intermetallic alloys \cite{Soma1, Soma2, Soma3}. RPES has been effectively used to get the details of the electronic structure near $E_F$ and its role on the superconducting properties in different classes of superconductors, namely, the iron Pnictides \cite{koi08, koi10, yok12, suz13} , YBa$_2$Cu$_3$O$_{7-\delta}$ \cite{kan02}, ZrB$_2$ \cite{tsu03}, PrPt$_4$Ge$_{12}$ \cite{nak10}, and the heavy Fermion superconductors \cite{raj05}. Therefore, we expect that the RPES studies on the Mo$_{1-x}$Re$_x$ alloys may provide useful information of the valence band, which in turn will be helpful in understanding the correlation between the ETT and the superconducting properties. In the present study, we have therefore performed RPES experiments on the $\beta$-phase Mo$_{1-x}$Re$_x$ alloys with $x > $ 0.1 and have compared these results with the results on elemental Mo. Our analysis reveals that the substitution of the Re in Mo increases the DOS at $E_F$ which then develops Re 5$d$ like character. These enhanced DOS at $E_F$, enhances the interband $s$-$d$ interaction in these alloys with increasing $x$, which in turn influences the nature of superconductivity.

\section{Experimental and Details of the band structural calculations}

Polycrystalline samples of Mo$_{1-x}$Re$_{x}$, where $x$ = 0,  0.15, 0.20, 0.25 and 0.40 were prepared by melting 99.95+ \% purity constituent elements in an arc-furnace under 99.999 \% Ar atmosphere. The samples were flipped and re-melted six times to improve the homogeneity. The X-ray diffraction study of these alloys shows that the samples have formed in the body centered cubic (bcc) phase (space group: Im$\bar{3}$m) \cite{shy15a, shy15b, shy15c}.

The RPES measurements on the above samples were performed using the angle-integrated photoemission beamline of the Indus-1 synchrotron radiation source \cite{cha02}. The valence band (VB) photoemission spectra were recorded in 23-70 eV photon energy range.  The details of the configuration of the present system and the experimental conditions are available elsewhere \cite{som4}. The spectra were normalized by the photon flux estimated from the photocurrent obtained from the post mirror of the beam line. Clean sample surface was obtained by sputtering the sample in-situ. Surface cleanliness was confirmed by the absence of the oxygen and the carbon contribution in the core level peaks of Mo and Re. The experimental resolution was estimated to be from 0.3-0.4 eV in the present photon energy range. The X-ray photoelectron spectra (XPS) were measured using Mg K$_{\alpha}$ X-ray source (DAR400, Omicron).

The ab-initio electronic structure calculations were performed using the spin polarized Korringa-Kohn-Rostoker (KKR) method \cite{akai}. The effect of doping was considered under the coherent potential approximation. The exchange correlation functional developed by Vosko Wilk and Nusair was used for the calculation \cite{vos80}. The number of $k$-points used in the irreducible part of the Brillouin zone is 72. The muffin-tin radii for Mo and Re atoms used in the calculations are same and equal to 2.576 bohr. For the angular momentum expansion, we have considered $l_{max}$ =2 for each atom. The potential convergence criterion was set to 10$^{-6}$.

\section{Results and discussion}

\begin{figure}
\centering
\epsfxsize=80mm
%\vskip -0.8cm
\epsffile{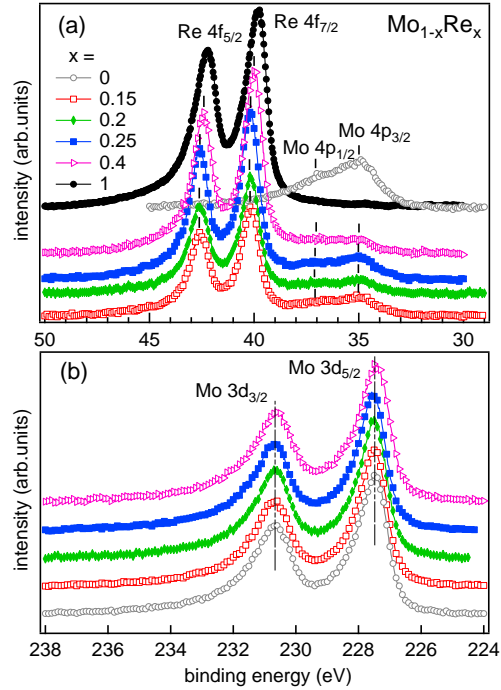}
\caption {Core level spectra of the Mo$_{1-x}$Re$_{x}$ alloys, where the panel ($a$) shows the Re 4$f$ and Mo 4$p$ states and the panel ($b$) shows the Mo 3$d$ states. The positions of the Re $4f$ corelevels in alloys which are marked as tick in ($a$) are found to shift with respect to that of elemental Re. The positions of the Mo $4p$ and Mo $3d$ corelevels in alloys which are marked as tick in ($a$) and ($b$) respectively remain at the same position as that of elemental Mo. The compositions $x$ = 0 and 1 represent elemental Mo and Re respectively.} 
\label{fig1}
%\vskip -0.5cm
\end{figure}

Fig. 1 ($a$), and ($b$) show the XPS data corresponding to the Mo $4p$, Re $4f$, and Mo $3d$ core-levels in the Mo$_{1-x}$Re$_{x}$ alloys. These core level spectra are also compared with those of elemental Mo and Re. The inelastic background has been subtracted using standard Tougaard method \cite{Tougaard89}. In Fig. 1($a$), the $4p_{1/2}$ and $4p_{3/2}$ peaks for elemental Mo appear at 37~eV and 35~eV binding energy (BE), respectively, showing a spin orbit splitting of 2~eV. On the other hand, the Re $4f_{5/2}$ and $4f_{7/2}$ peaks appear at 42.2~eV and 39.8~eV respectively, showing a spin orbit splitting of 2.4~eV. The interesting observation here is that the BE positions of the Mo $4p_{1/2}$ and $4p_{3/2}$ core levels in the Mo$_{1-x}$Re$_{x}$ alloys appear at the same BE positions as that of elemental Mo (Fig 1($a$)). However, the Re $4f_{5/2}$ and $4f_{7/2}$ show a shift towards higher binding energy with decreasing $x$ as compared to elemental Re. Within the instrumental resolution, the shift observed for $x$=~0.4, is 0.25~eV. For $x$=~0.15, 0.2 and 0.25, the shift is about 0.4~eV towards higher binding energy. In a binary disordered metallic alloy, such a shift appears due to the difference in the local chemical environment and/or the charge redistribution, which is related to the hybridization between the valence electron states of the system \cite{olo01}. In Fig.~1($b$), the BE peak positions for the Mo 3$d$ core levels (Mo $3d_{5/2}$ = 227.8 eV \& Mo $3d_{3/2}$ = 231.0 eV) in the Mo$_{1-x}$Re$_{x}$ alloys are found at the same BE positions (within the experimental resolution) as those reported for elemental Mo \cite{wer83}. Similar to Mo 3$d$ core levels (Fig. 1($b$)), the peak positions for Re 4$d$ core levels in the Mo$_{1-x}$Re$_{x}$ alloys (not shown here), are also found to be same as of elemental Re.

\begin{figure}
\centering
\epsfxsize=70mm
%\vskip -0.8cm
\epsffile{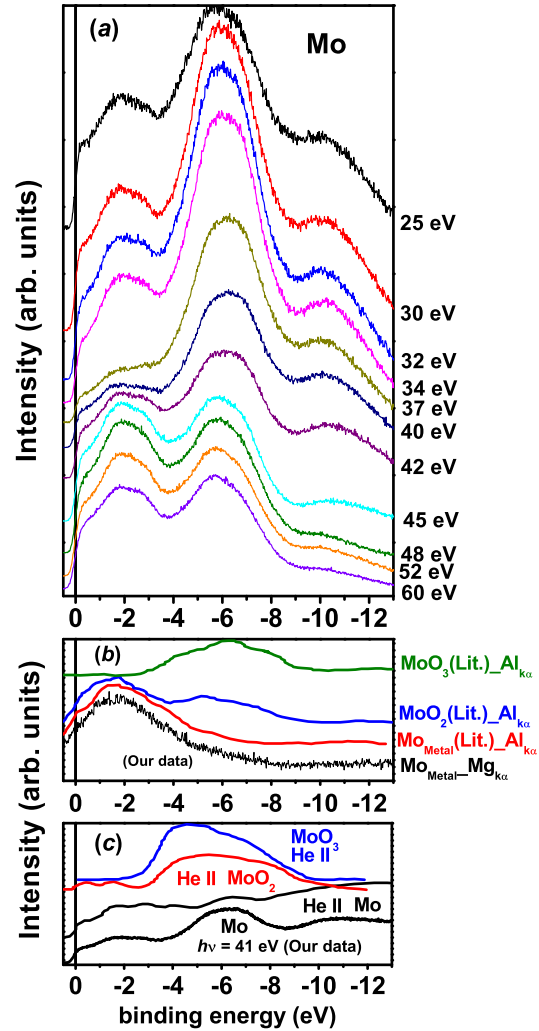}
\caption {($a$) Valence band photoelectron spectra (binding energy vs. intensity) for pure Mo in the photon energy range 23 eV to 60 eV. ($b$) Valence band XPS data for Mo-oxides and metallic Mo (data extracted from Ref. \cite{wer83}) in comparison with our experimental data for metallic Mo. ($c$) Valence band UPS data (He II source) for Mo-oxides and metallic Mo (data extracted from Ref. \cite{tok01,sch78}) compared with our experimental data for metallic Mo at 41 eV photon energy. Panels ($b$) and ($c$) clearly show the absence of oxygen in this system.} 
\label{fig2}
%\vskip -0.5cm
\end{figure} 

The VB spectra for pure Mo in the energy range 23 to 60~eV are shown in Fig. 2 ($a$). The background obtained by the Tougaard procedure \cite{Tougaard89} has been subtracted from the raw data. In Fig. 2 ($b$), we compare the present VB-XPS of elemental Mo measured using Mg K$\alpha$ radiation with the XPS results reported in literature for Mo-oxides and elemental Mo where the measurements were done using Al K$\alpha$ radiation \cite{wer83}. From Fig. 2 ($b$), it is clear that the VB-XPS of elemental Mo is in fairly good agreement with that of the literature and is quite different from that of oxides of Mo. Therefore, we believe that there is no oxygen in the bulk of the samples. Since, the photons with energies such as 20-30~eV are more of surface sensitive nature, we compare in Fig. 2($c$) the photoelectron spectrum (PES) of the VB of Mo obtained using photons of 41~eV energy with the reported PES of the VB of elemental Mo and Mo-oxides obtained using He-II source \cite{tok01,sch78}. The features corresponding to the VB of elemental Mo are distinctly different from that of Mo-oxides wherein broad hump like features are observed centered around -5~eV below $E_F$. Therefore, we believe that the surface of present samples under study is not contaminated by oxygen. The PES spectra of the VB of Mo measured at 41~eV photon energy showing fine features up to about -4~eV below $E_F$, and then a broad hump ranging from -4~eV to -8~eV which is followed by another hump like feature centered around -10~eV below $E_F$, are in agreement with that measured \cite{sch78} at 40.8~eV photon energy (He-II source) as well as other measurements reported in literature \cite{mik81}.

The PES spectra of Mo was taken up to BE = -14~eV below $E_F$, consists of five features viz. one feature very close to $E_F$, two features closely spaced around -2 to -3~eV, a broad peak at -6~eV and a broad hump centered around -10~eV below $E_F$. All the peaks are quite broad due to the large dispersion of the bands \cite{oka13}. A systematic shift of the positions of these peaks are also observed when the incident photon energy is varied. This may be related to the dispersion of the bands along the k$_z$ direction of the Brillouin zone. These results are consistent with many theoretical and experimental studies performed on Mo (110), (011) and (112) face single crystals \cite{shan78, shan77, yak97, jeo88, cin76, jeo89, yak01}. Since, the low energy photons are surface sensitive, the surface density of states also influence the PES. Weng et al., have shown that there are two peaks centered around -0.7~eV and -3~eV below E$_F$ in the density of states corresponding to the Mo surface \cite{shan78}. We have also observed the corresponding features in the PES, which is probably the admixture of the bulk and surface states. However, we have also observed other features at -2~eV, -6~eV and -10~eV which are mainly from the bulk of the sample \cite{shan78}. The resonant enhancement of the PES of the VB is observed as a function of photon energy which will be discussed later in detail.

\begin{figure}
\centering
\epsfxsize=100mm
%\vskip -0.8cm
\epsffile{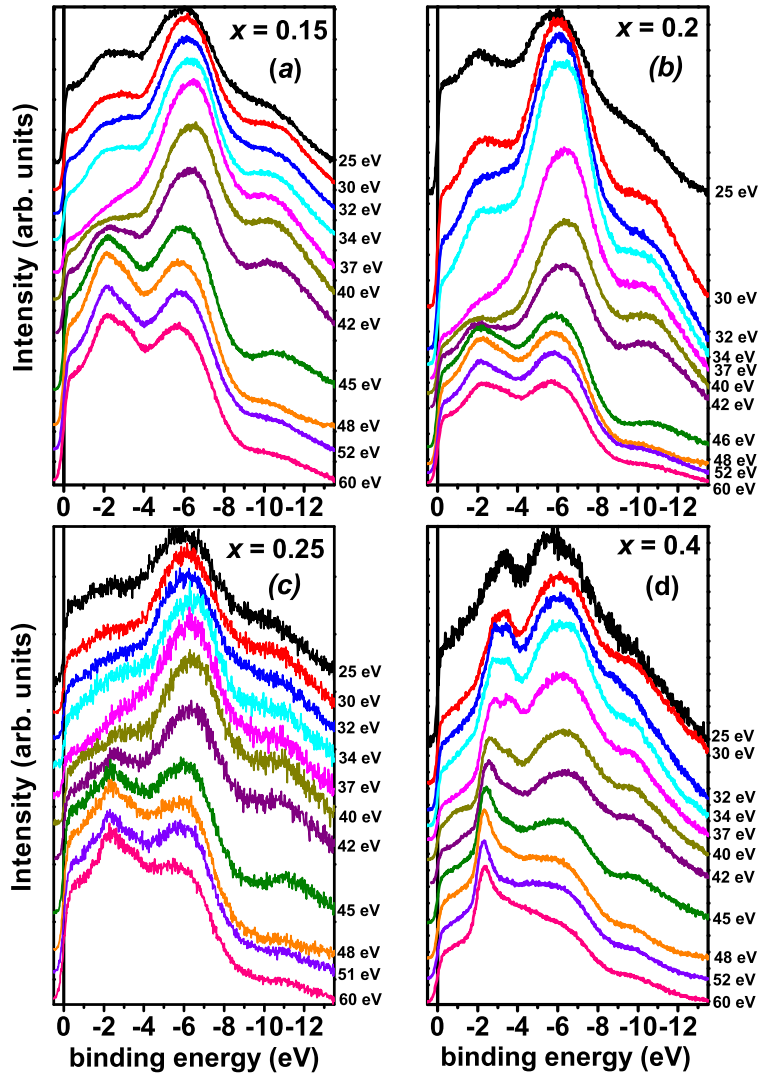}
\caption {Valence band photoemission spectra of Mo$_{1-x}$Re$_{x}$ alloys for $x$~=~0.15, 0.2, 0.25 and 0.4 samples, plotted at some selected photon energies across Mo 4$p$-4$d$ and Re 5$p$-5$d$ resonances.} 
\label{fig3}
%\vskip -0.5cm
\end{figure}

A systematic evolution of the VB photoemission spectra of the Mo$_{1-x}$Re$_{x}$ alloys with $x$~= 0.15, 0.2, 0.25 and 0.40 measured at the selected photon energies ranging from 25~eV to 60~eV is shown in Fig. 3. In comparison with Mo, the intensity of the BE feature in the Mo$_{1-x}$Re$_{x}$ alloys at -6~eV decreases monotonically as a function of photon energy and this reduction of intensity is quite significant for the alloys with higher Re content. The features around BE = -2~eV become sharp when Re alloyed with the Mo, although the peak positions of these features do not seem to depend on the photon energy. However, when Re is alloyed with Mo, the features in the PES depends on the photon energy quite significantly and the average shift in the peak positions as a function of the photon energy increases with increasing Re content. This is due to the enhancement in the dispersion of the bands along the k$_z$ direction of the Brillouin zone when Re is alloyed with Mo \cite{oka13}.

\begin{figure}
\centering
\epsfxsize=90mm
%\vskip -0.8cm
\epsffile{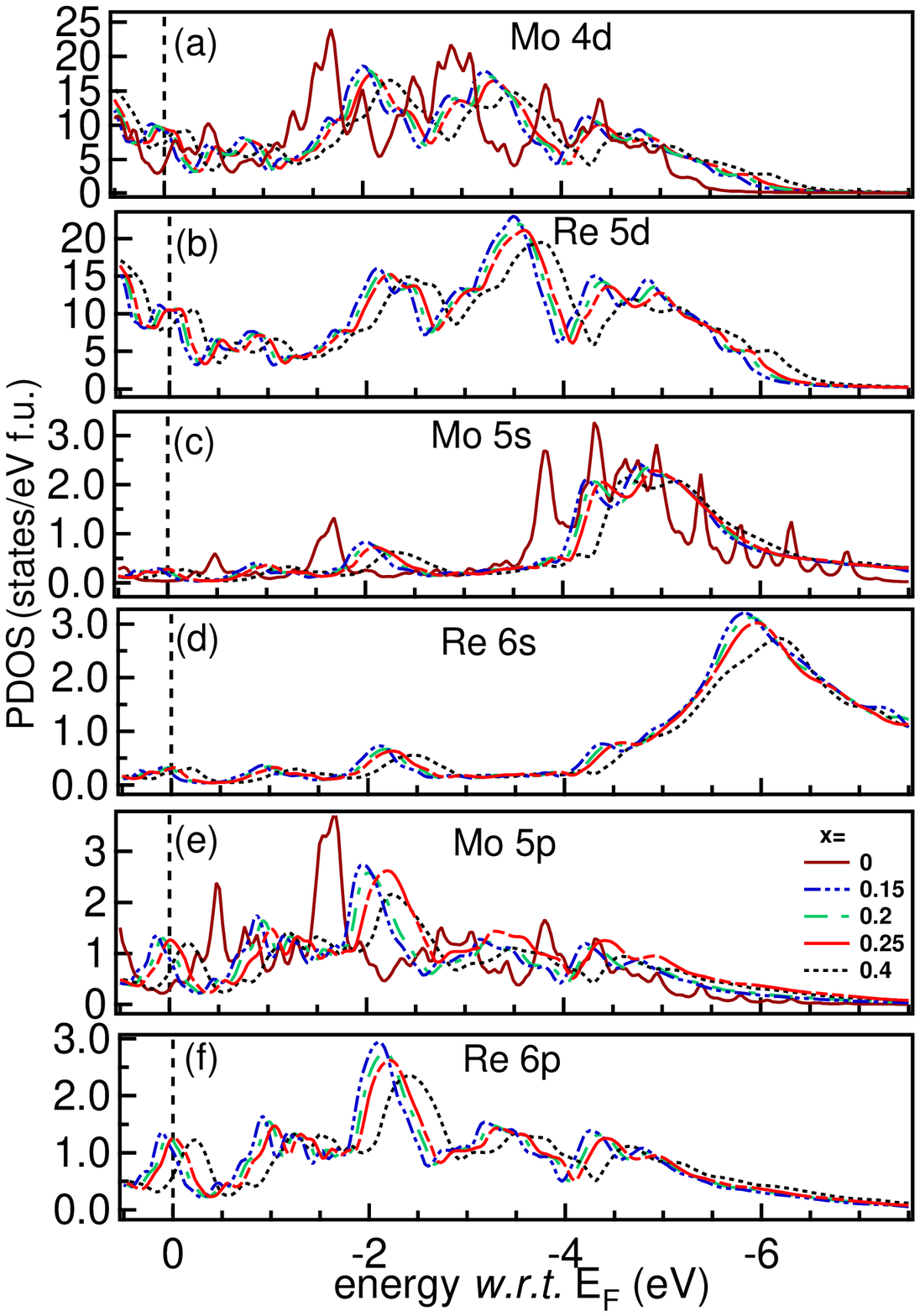}
\caption {Partial DOS showing Mo $4d$, Re $5d$, Mo $5s$, Mo $5p$, Re $6s$, $6p$ states in ($a$) to ($f$) respectively.} 
\label{fig8}
%\vskip -0.5cm
\end{figure}

In order to know the contributions of different orbitals to the PES, we plot in Fig. 4, the partial density of states (PDOS) of Re and Mo in the Mo$_{1-x}$Re$_{x}$ alloys. The $d$ and $p$ states of both Mo and Re lie in the BE range $E_F$ to -6~eV, and $s$ states of both Mo and Re lie below -4~eV in the higher BE side. The 4$d$ states of Mo show three peaks around BE = -2, -3 and -4~eV below $E_F$, whereas the 5$d$ states of Re show three peaks around BE = -2.5, -3.5 and -5~eV below $E_F$. At the Fermi level, a dip is observed in the Mo $4d$ PDOS. When Re is alloyed with Mo, the contributions of both Mo and Re $d$ states increase with the increase in $x$ up to $x$ = 0.25. In the case of $x$ = 0.40, however, the $d$ DOS is reduced once again. The Mo 5$s$ states lie around -5~eV below $E_F$ whereas the Re 6$s$ states lie around -6~eV below $E_F$. However, the contribution of the $s$ states of both Mo and Re in the energy range $E_F$ to -4~eV below $E_F$ is quite small but finite. The Mo 5$p$ and Re 6$p$ states are evenly distributed from $E_F$ to -4~eV below $E_F$ but rather quite less in number as compared to the $d$ states. Since the Re (5$d^5$6$s^2$) has one more valence electron per atom than Mo (4$d^5$5$s^1$), the PDOS of both Mo and Re shift towards higher binding energy with the increase in Re concentration. Therefore, features up to about -4~eV below $E_F$ in the PES is dominated by the $d$ like states and features below -4~eV towards higher binding energy represents the $s$ like states.

\begin{figure}
\centering
\epsfxsize=80mm
%\vskip -0.8cm
\epsffile{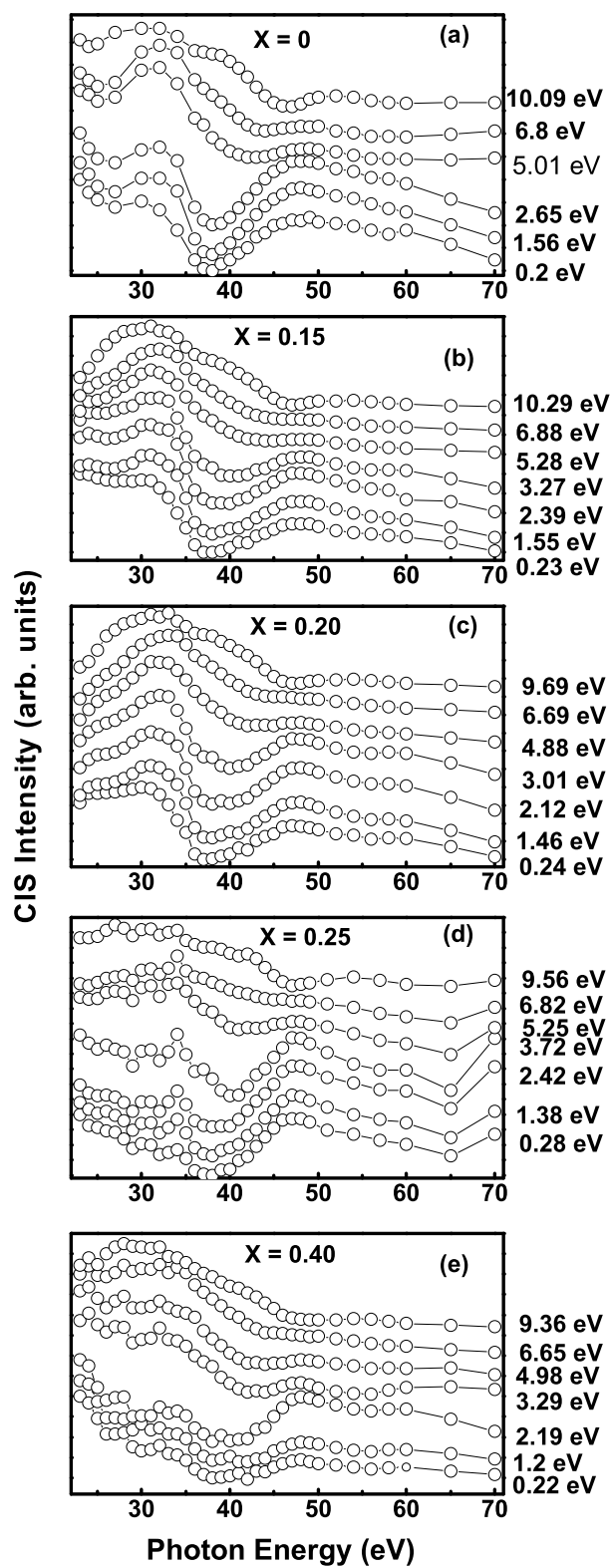}
\caption {The constant initial state (CIS) of the valence band features for Mo$_{1-x}$Re$_{x}$ alloys (where $x$=~0, 0.15, 0.2, 0.25 and 0.4) as a function of incident photon energy. The valence band features at different binding energy positions are staggered for the clarity of presentation.} 
\label{fig4}
%\vskip -0.5cm
\end{figure}

Fig. 2 and Fig. 3 show the resonance enhancement of the features in the PES measured using photons with different energies. In all the samples, the feature around BE = -6~eV below $E_F$ resonates around 35~eV photon energy, whereas the feature around BE = -2~eV below $E_F$ resonates around 45~eV photon energy. A clear picture can be obtained by plotting the constant initial state (CIS) intensities of the VB features as a function of photon energies as shown in Fig. 5. The CIS plots of all the samples show two resonances with the peak positions at around 35~eV and 45~eV incident photon energy. 
In the case of Mo, the resonance corresponding to the incident photon energy of around 35~eV is observed over the entire valance band, whereas, the resonance corresponding to the incident photon energy of around 45~eV is observed only below the 5~eV binding energy of the V B. From the Fig. 1, it is clear that the resonance of the VB occurred across the Mo 4$p$ threshold. The band structure calculation (Fig. 4) suggests that the features of the PES in the BE range from $E_F$ to -5~eV below $E_F$ are mainly derived from $d$ like states where the features of the PES in BE range from -5~eV below $E_F$ to -8~eV below $E_F$ are mainly derived from $s$ like states. Therefore, the resonance around 35~eV photon energy is from Mo 4$p$-5$s$ transition via: \\ 1)Direct photoemission: \\$4p^64d^55s^1 + h\nu~\rightarrow~ 4p^64d^55s^0 + e^-$ and \\ 2) Auger emission: \\$4p^64d^55s^1 + h\nu~\rightarrow~ 4p^54d^55s^2~\rightarrow~4p^64d^55s^0 + e^-$.\\ whereas the resonance around 45~eV photon energy is from Mo 4$p$-4$d$ transition via: \\ 1)Direct photoemission: \\$4p^64d^55s^1 + h\nu~\rightarrow~ 4p^64d^45s^1 + e^-$ and \\ 2) Auger emission: \\$4p^64d^55s^1 + h\nu~\rightarrow~ 4p^54d^65s^1~\rightarrow~4p^64d^45s^1 + e^-$.\\ \\ The angle resolved photoemission spectroscopy (ARPES) study on the Mo(100) surface at various photon energies by Weng et al., revealed four resonances at about 15~eV, 30~eV, 38~eV and 45~eV photon energies \cite{shan78, shan77}. They have argued that the resonances of the features observed at B.E. = -0.3~eV and -3.3~eV below $E_F$ for the photon energies of about 15~eV, 30~eV, and 38~eV are related to surface states. In the present case, both the resonances are observed in the entire VB indicating that these resonances are related to bulk states. However, the resonances around BE = -0.3~eV and -3.3~eV below $E_F$ are to be used with care as they might be influenced by the surface states as well. It is worth mentioning that the surface resonance feature is extremely sensitive to the surface contamination \cite{shan78, shan77}. Both the polycrystalline nature of the samples and the appearance of these resonances in the alloys suggest that these resonances are mainly related to the bulk states. Similar to Mo, we have observed two resonances in the Mo$_{1-x}$Re$_{x}$ alloys. As the Re content is increased (from $x$ = 0), the resonance corresponding to the incident photon energy of around 35~eV diminishes below 5~eV and become non existent for x $\geq$ 0.25. The resonance corresponding to the incident photon energy of around 45~eV shifts to higher incident photon energy when Re is alloyed with Mo. This is due to the fact that in these alloys, an additional contribution from Re $5p$-$5d$ transition (46~eV) is expected for the resonance at around 45~eV photon energy. This additional contribution originates from the Re 5$p$-5$d$ transition (46~eV) via: \\ 1) Direct photoemission:\\$5p^65d^56s^2 + h\nu~\rightarrow~ 5p^65d^46s^2 + e^-$ and \\2) Auger emission: \\$5p^65d^56s^2 + h\nu~\rightarrow~ 5p^55d^66s^2~\rightarrow~5p^65d^46s^2 + e^-$.\\ \\ 

\begin{figure}
\centering
\epsfxsize=100mm
%\vskip -0.8cm
\epsffile{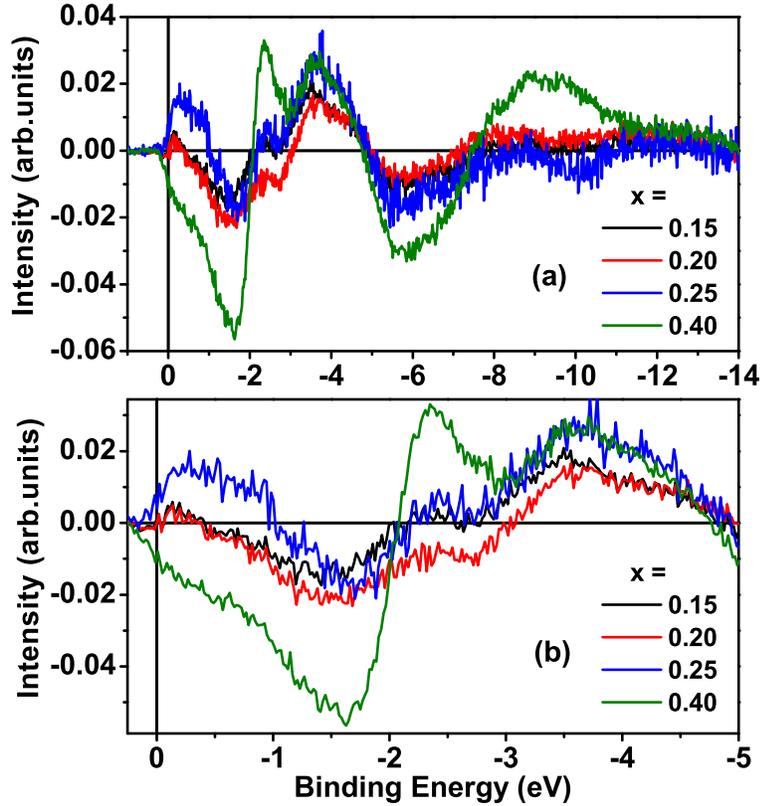}
\caption {Mo and Re contribution extracted by taking the difference between photoemission spectra of Mo$_{1-x}$Re$_{x}$ alloys and pure Mo at 47~eV photon energy. The details are explained in the text.} 
\label{fig7}
%\vskip -0.5cm
\end{figure}

In order to know the effect of alloying Re with Mo on the PES we plot in Fig. 6, the difference between the spectra for the Mo$_{1-x}$Re$_{x}$ alloys and pure Mo at 47~eV photon energy. The positive contribution implies the enhancement in the intensity of the PES features with respect to that of Mo introduced by the Re states. On the other hand, the negative contribution corresponds to the loss of Mo states due to alloying and/or the absence of Re states at the corresponding BE. The expanded portion of the difference spectra up to -5~eV below $E_F$ is shown in panel ($b$) of Fig. 6. This figure suggests that the contribution of DOS from the Re $d$ states increases near $E_F$ with the increase in $x$ up to 0.25. The width of the Re $d$ states near $E_F$ also increases with the increase in $x$ up to 0.25. For $x$~=0.15  and 0.20, the Re $d$ states are observed up to about -0.5~eV below $E_F$, whereas for $x$ = 0.25, the Re $d$ states are observed up to about -1~eV below $E_F$. On the other hand, the difference spectra for $x$ = 0.40 is quite different from the rest of the alloys. In this case, the contribution to the difference spectra around $E_F$ is negative which indicates the loss of DOS in spite of the higher Re content. These results are in agreement with the DOS calculation for $x$ = 0.4 (Fig. 4) which also indicate a decrease in the $d$ partial DOS of Mo and Re as compared to the Mo$_{1-x}$Re$_{x}$ alloys. Additionally, in the higher binding energy side from -2 to -5~eV, there is significant contribution from the Re $d$ states.

\begin{figure}
\centering
\epsfxsize=120mm
%\vskip -0.8cm
\epsffile{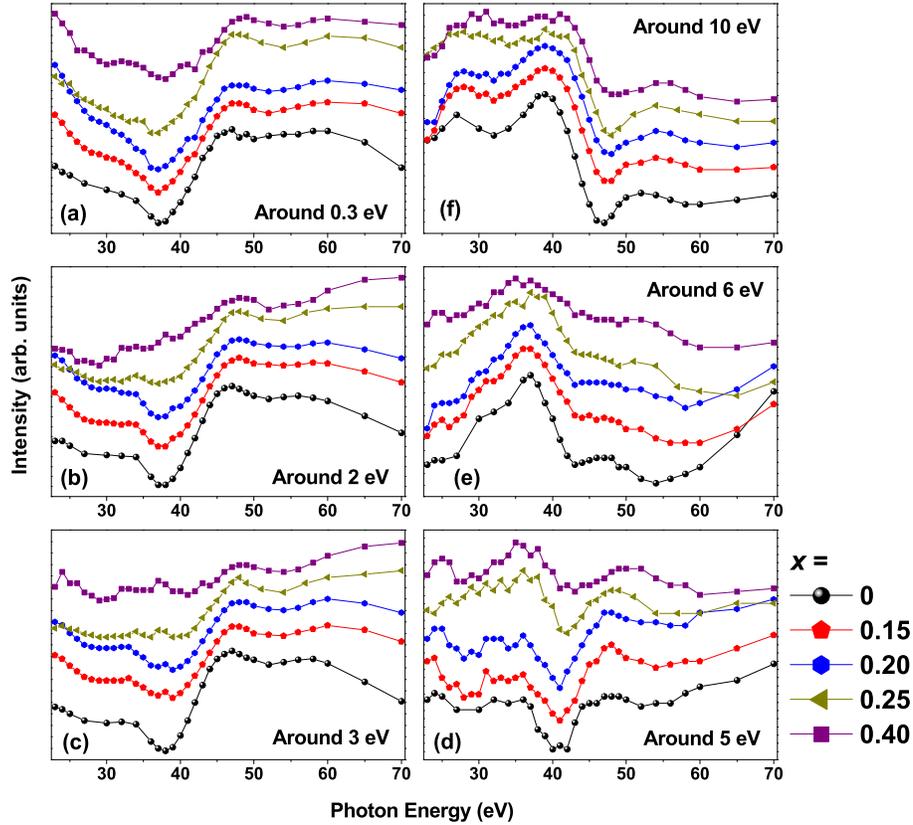}
\caption {Comparison of the area normalized intensities as a function of incident photon energy at different binding energies of the valance band (indicated in the figure) for Mo$_{1-x}$Re$_{x}$ alloys with $x$=~0, 0.15, 0.2, 0.25 and 0.4.}
\label{fig4}
%\vskip -0.5cm
\end{figure}

The ARPES study performed by Okada et al. \cite{oka13} shows that an extra band appears at the Fermi surface of the Mo$_{1-x}$Re$_{x}$ alloys above 10 at.\% of Re as a result of ETT in these alloys. When these results are compared with the difference spectra shown in Fig. 6, it is clear that the extra band that appeared at the Fermi surface above 10 at.\% of Re has Re $d$ like character. Recently we have shown that this extra band contribute distinctly to the superconductivity \cite{shy15c}. The temperature dependence of the normal state resistivity of these alloys indicated that the phonon-assisted $s$-$d$ scattering in Mo is weak as compared to the intra-band $s$-$s$ scattering, and that the $s$-$d$ scattering is enhanced with the increase in $x$ above 10 at.\% of Re \cite{shy15a}. In the case of $s$-$d$ scattering, the resistivity is proportional to $N_d(E_F)/N_s(E_F)$ \cite{sha10} where $N_d(E_F)$ and $N_s(E_F)$ are the partial DOS of the $d$ and $s$ states at the Fermi level. In order to get further insight into it, we have compared in Fig. 7, the area normalized intensities as a function of incident photon energy at different binding energies of the valance band for the present Mo$_{1-x}$Re$_{x}$ alloys.  The figure reveals that the resonance corresponding to the incident photon energy of around 35~eV is present in Mo over the entire valance band. However, when Re is alloyed with Mo, the resonance corresponding to the incident photon energy of around 35~eV diminishes below BE = -5~eV and vanishes for $x \geq$ 0.25, whereas for BE $\geq$ -5~eV, this resonance is present in all the alloy compositions. Note that the $d$ states are present in the range E$_F$ to -5~eV below E$_F$ along with a small contribution from the $s$ states. This indicates that in Mo, the photon energy corresponding to the resonance of PES involving the $s$ states of the VB is distinctly different from that involving the $d$ states implying negligible $s$-$d$ interaction. The present study reveals that there is an enhancement of the $d$ DOS in the vicinity of the Fermi surface when Re is alloyed with Mo which results in the enhancement of $s$-$d$ scattering in the alloy compositions. Then, the final state 4$p^5$4$d^5$5$s^2$ formed due to the Auger process at about the incident photon energy of 35~eV can return to the original state via $s$-$d$ scattering. This results in the diminishing of the resonant features below BE = -5~eV for the incident photon energy of 35~eV when Re is alloyed with Mo up to $x$ = 0.20 and its subsequent disappearance for $x \geq$ 0.25. We have previously shown that the enhancement of superconducting transition temperature $T_c$ in these alloys is related to the $s$-$d$ scattering. A special mention is needed here for the Mo$_{0.60}$Re$_{0.40}$ alloy, as the present photoemission as well as band structure calculations showed the decrease in the density of states at the $E_F$ as compared to $x$ = 0.25, but, the $T_c$ enhanced to 12.6~K compared to 9.6~K  for $x$ = 0.25. This can be understood in terms of the dispersion of bands along the H-N direction of the Brillouin zone. The momentum $k$ separation between the band that appeared in the H-N direction after alloying and the other band with Mo-like character reduces with the increase in $x$. Therefore, for the alloys with higher concentrations of Re, the low energy phonons (small $k$) can help in $s$-$d$ scattering. Thus, the multi-band effect is smeared out and the $T_c$ is enhanced in the Mo$_{0.60}$Re$_{0.40}$ alloy.                        

\section{Conclusion}

We have performed resonant photoemission spectroscopy experiments on the Mo$_{1-x}$Re$_{x}$ alloys with $x$ = 0, 0.15, 0.2, 0.25, and 0.4 in the photon energy range 23-70~eV. A resonance enhancement of intensity of the photo-electron spectra corresponding to Mo $4p$-$4d$ and/or Re $5p$-$5d$ transitions at about 45~eV incident photon energy is observed in all the alloys. We have observed another resonance in Mo and low $x$ alloys at about 35~eV incident photon energy. Our analysis reveals that the resonance at 35~eV is related to the Mo $4p$-$5s$ transition. The two separate resonances at $E_F$ for pure Mo and lower Re content alloys indicate the lack of $s$-$d$ interaction in these alloys. The PES of all the alloys are in agreement with the theoretical DOS estimated from band structure calculations. The analysis suggests that experimentally observed enhancement of the $d$ like states at the Fermi level and its dispersion upon alloying Re with Mo governs the physical properties of the both normal and superconducting states. 

\section{Acknowledgement}
We thank Shri R. K. Meena for sample preparations and Shri A. D. Wadikar for his help in the photoelectron spectroscopy experiments.

\section*{References}

%\vskip -0.2cm
%\newpage

%\bibliographystyle{iopart-num}
%\bibliography{PES-MS}
\providecommand{\newblock}{}

\end{document}